\documentclass[12pt,prd,showkeys,amsmath,amssymb,nofootinbib]{revtex4-1}
\usepackage{amsmath,amsfonts,amsthm,amscd,amssymb,latexsym}
\usepackage{bm}
\usepackage{dcolumn}
\usepackage{graphicx}
\usepackage{epstopdf}
\usepackage{color}
\usepackage{epsf}
\usepackage{epsfig}
\usepackage{graphicx, epic, eepic, color}
\usepackage[colorlinks=true,urlcolor=blue,linkcolor=blue,citecolor=blue]{hyperref}

\begin{document}

\title{Spin-of-Light Gyroscope and the Spin-Rotation Coupling}

\author{Bahram \surname{Mashhoon}$^{1,2,3}$}
\email{mashhoonb@missouri.edu}
\author{Yuri N. Obukhov$^4$}
\email{obukhov@ibrae.ac.ru}

\affiliation{$^1$School of Astronomy, Institute for Research in Fundamental Sciences (IPM), Tehran 19395-5531, Iran\\
$^2$Department of Physics, Sharif University of Technology, Tehran 11365-9161, Iran\\
$^3$Department of Physics and Astronomy, University of Missouri, Columbia, Missouri 65211, USA\\
$^4$Theoretical Physics Laboratory, Nuclear Safety Institute, \\
Russian Academy of Sciences, B. Tulskaya 52, 115191 Moscow, Russia
}

\date{\today}

\begin{abstract}
We discuss the coupling of photon spin with rotation in connection with a recent proposal of Fedderke \emph{et al.} [arXiv:2406.16178] regarding a precision gyroscope based on the intrinsic spin of light. To this end, we analyze the propagation of electromagnetic radiation in a physical system that uniformly rotates about the direction of wave propagation in the presence of an ambient medium. Finally, we consider the possibility of using this type of spin-of-light gyroscope to measure gravitomagnetic fields.  
\end{abstract}


\maketitle

\section{Introduction}

In a recent paper, Fedderke \emph{et al.}~\cite{Fedderke:2024ncj} have proposed a gyroscope in which light primarily travels parallel to the axis of uniform rotation and the design of the device is based on the coupling of photon spin with rotation. The main purpose of this paper is to elucidate the main theoretical issue upon which the new gyroscope is based. 

Let us consider an observer that rotates uniformly with angular velocity $\bm{\Omega} = \Omega\,\bm{e}_z$ and velocity $\bm{v} = \bm{\Omega}\times\bm{r}$ about the $z$ axis in a global inertial frame of reference. Here, $\bm{e}_z$ is a unit vector along the $z$ axis and $\bm{r}$ is the observer's position in space. The observer measures the frequency of an incident  plane monochromatic electromagnetic wave of frequency $\omega_0$ and wave vector $\bm{k}_0$. In the WKB approximation ($\omega_0 \gg \Omega$), the measured frequency of the wave is given by the relativistic Doppler formula $\omega_D = \gamma (\omega_0 -  \bm{v} \cdot \bm{k}_0)$; hence,
\begin{equation}\label{1}
\omega_D = \gamma (\omega_0 - \bm{\Omega} \cdot \bm{\ell})\,,
\end{equation}
where $\bm{\ell} = \bm{r} \times \bm{k}_0$. In terms of energy $E_0 = \hbar\,\omega_0$, 
\begin{equation}\label{2}
E_D = \gamma (E_0 - \bm{\Omega}\cdot\bm{L})\,, \qquad \bm{L} = \hbar\,\bm{\ell}\,,
\end{equation}
where $\bm{L}$ is the orbital angular momentum of the light ray. However, the generator of rotations is the total angular momentum $\bm{J} = \bm{L} + \bm{S}$, where $\bm{S} = \pm\hbar\,\bm{k}_0/|\bm{k}_0|$ represents the photon spin. Therefore, we have the more general result
\begin{equation}\label{3}
E = \gamma (E_0 - \bm{\Omega} \cdot \bm{L} - \bm{\Omega} \cdot \bm{S})\,,
\end{equation}
which exhibits the spin-rotation coupling~\cite{BMB}.

In interferometry, the coupling of orbital angular momentum with rotation appears as the Sagnac effect~\cite{Sagnac}. The Sagnac phase shift, 
\begin{equation}\label{4}
\Delta \Phi_{\rm Sagnac} = \frac{4\omega_0}{c^2}\,\bm{\Omega} \cdot \bm{A}\,
\end{equation}
is proportional to the scalar product of the angular velocity of rotation and the area $\bm{A}$ of the Sagnac interferometer and is independent of the motion of the observer. The Sagnac effect was employed by Michelson \emph{et al.}~\cite{MGP} to measure the angular velocity of rotation of the Earth. In interferometry with radiation of particles with rest mass $m > 0$, we must replace $\omega_0$ in the Sagnac formula (\ref{4}) with the de Broglie frequency of the matter waves~\cite{Page:1975zz}. For further work on the Sagnac effect, see~\cite{Dresden:1979sv, WSC, Sakurai:1980te, RKWHB, HN, Anderson:1994hx, Stedman:1997wm, Mashhoon:1998dm}. Interesting new studies regarding the photon and neutron Sagnac effects  are contained in~\cite{KWH, Geerits:2024jdt}.  

It follows from Sagnac's formula (\ref{4}) that the Sagnac effect vanishes if the axis of rotation lies in the plane of the interferometer. In this situation, it is in principle possible to have an inertial sensing device that is purely based on the two helicities of the electromagnetic radiation~\cite{BM1989}. In general, both Sagnac and spin effects could be present; in the optical regime, the spin effect is smaller than the Sagnac effect by the ratio of the wavelength of light to the linear dimension of the interferometer. To distinguish the spin effect from the Sagnac effect, we note that the spin effect is independent of the wavelength and changes sign with the reversal of the helicity of the radiation~\cite{BM1989}.

\section{Helicity-rotation coupling} 

The phenomenon of spin-rotation coupling is based upon the inertia of intrinsic spin. Specifically,  the intrinsic spin $\bm{S}$ of a particle couples to the rotation of a noninertial observer resulting in a Hamiltonian of the form
\begin{equation}\label{I1}
\mathcal{H}_{sr} =  -\bm{S} \cdot \bm{\Omega}\,,
\end{equation} 
where $\bm{\Omega}$ is the angular velocity of the observer's local spatial frame with respect to a nonrotating (i.e., Fermi-Walker) transported frame. We neglect the Lorentz factor in Eq.~(\ref{I1}), since all terms of second order in $\Omega$ will be ignored in our analysis. A comprehensive treatment of the spin-rotation coupling is contained in~\cite{BMB}; furthermore,  recent discussions of its observational basis in neutron physics can be found in~\cite{DSH, DDSH, DDKWLSH}, while its applications in spintronics are discussed in~\cite{Yu:2022vjn}.  Moreover, the connection between the spin-rotation coupling phenomena and the special theory of relativity has been elucidated in~\cite{Mashhoon:2024qwj}.

Let us imagine that a monochromatic plane electromagnetic wave has frequency $\omega_0$ and wave vector $\bm{k_0}$ in a global inertial frame with Cartesian coordinates $(t_0, x_0, y_0, z_0)$. A reference observer is at rest at the origin of spatial coordinates, but refers its measurements to axes that rotate with respect to Cartesian axes with constant angular speed $\Omega$ about the $z_0$ axis.  In the frame rotating with uniform frequency $\bm{\Omega} = \Omega\,\bm{e}_{z_0}$, the fiducial noninertial (i.e., rotating) observer  finds~\cite{BMB}
\begin{equation}\label{I2}
\omega = \omega_0 - \bm{H}_0 \cdot \bm{\Omega}\,, \qquad
\bm{H}_0 = \pm\,\hat{\bm{k}}_0 = \pm\, \frac{\bm{k}_0}{|\bm{k}_0|}\,,
\end{equation}
where $\bm{H}_0$ is the helicity vector and
\begin{equation}\label{I3}
\bm{k} = \bm{k}_0\,,
\end{equation} 
as a consequence of spin-rotation coupling. A detailed derivation of these results is contained in~\cite{BMB, Mashhoon:2002fq, Hauck:2003gy, Mashhoon:2024qwj}. A simple illustration of the helicity-rotation coupling when the radiation is incident along the rotation axis is presented in Appendix~\ref{appA}; on the other hand, the situation is more complicated for oblique incidence. In this case, a wave mechanical approach implies that the \emph{average} frequency of the incident radiation measured by the noninertial observer is given by Eq.~(\ref{I2}).  The same result follows from the eikonal approximation when $\omega \gg \Omega$~\cite{Mashhoon:2002fq}.

From $\omega_0 = c k_0$, $k_0 = |\bm{k}_0|$, in the inertial frame, we find the dispersion relation in the uniformly rotating frame, namely, 
\begin{equation}\label{I4}
\omega +  \bm{H} \cdot \bm{\Omega} = ck\,,
\end{equation} 
where  $k$ is the magnitude of the wave vector. Let us note that in the helicity vector the upper sign refers to positive helicity radiation while the lower sign refers to negative helicity radiation. In the uniformly rotating frame, the spacetime is stationary; hence, the frequency remains constant. Let $\omega$ be the  constant frequency of radiation for both helicity states; accordingly, Eq.~(\ref{I4}) implies that the wave vector must change in the rotating frame; therefore, 
\begin{equation}\label{I5}
\omega \pm \frac{\bm{k}^{\pm}}{|\bm{k}^{\pm}|} \cdot \bm{\Omega} = ck^{\pm}\,.
\end{equation} 
For propagation along the direction of uniform rotation, we find
\begin{equation}\label{I6}
ck^{\pm}  = \omega \pm \Omega\,.
\end{equation} 
This circumstance is entirely analogous to the situation of light propagating in the stationary spacetime of a rotating gravitational source that results in the Skrotskii effect, namely, the rotation of the plane of linear polarization also known as the gravitational Faraday effect; see Appendix~\ref{appB} and the references cited therein. 

The result~(\ref{I6}) from the spin-rotation coupling agrees with Eq. (3) of Fedderke \emph{et al.}~\cite{Fedderke:2024ncj} if positive-helicity radiation corresponds to their left circularly polarized light and if all their considerations do indeed take place in the rotating frame.  

The generalization of Eq.~(\ref{I6}) to include the possible influence of an ambient medium is the subject of the next section. In this paper, Greek indices label four-dimensional spacetime components $\mu,\nu,\dots = 0,1,2,3$, while Latin indices are used for three-dimensional spatial objects $a,b,\dots = 1,2,3$, anholonomic frame indices are marked by hats; moreover, the signature of the spacetime metric is +2.

\section{Rotating frame as an optical medium}\label{frame}

Maxwell's equations can be written in terms of the electromagnetic field tensor $F_{\mu \nu}$ and electromagnetic excitation\footnote{The terminology goes back to Mie and Sommerfeld \cite{Mie,Sommerfeld}.} tensor density ${\mathcal H}^{\mu \nu}$ as
\begin{equation}\label{J1}
F_{[\mu \nu , \rho]}  = 0\,, \qquad {\mathcal H}^{\mu \nu}{}_{,\nu} = 0\,, 
\end{equation} 
in the absence of sources~\cite{HeOb,Obukhov:2021ayi}. The energy-momentum tensor of the electromagnetic field is of second order in $F_{\mu \nu}$; therefore, to linear order in $F_{\mu \nu}$ the back reaction of the electromagnetic field on the background spacetime metric may be neglected. In the local coordinates\footnote{We employ $x^0 = t$ throughout this section; therefore, the Minkowski metric tensor is given by $\eta_{\mu\nu} = {\rm diag}(-c^2, 1, 1, 1)$.} $x^\mu = (t, x^a)$, the system of differential equations (\ref{J1}) can be viewed as electrodynamics in a hypothetical optical medium that occupies the space. With the natural identifications
\begin{eqnarray}
{E}_a = F_{a0}\,,\quad {B}^1 = F_{23}\,,\quad {B}^2 = F_{31}\,,
\quad {B}^3 = F_{12}\,,\label{EB}\\
{D}^a = {\mathcal H}^{0a}\,,\quad {H}_1 = {\mathcal H}^{23}\,,\quad {H}_2 = {\mathcal H}^{31}\,,
\quad {H}_3 = {\mathcal H}^{12},\label{DH}
\end{eqnarray} 
we recast (\ref{J1}) into the set of standard Maxwell's equations in a medium \cite{Obukhov:2021ayi} in terms of the fields $(\bm{E}, \bm{B})$ and excitations $(\bm{D}, \bm{H})$. One can check that 3-component fields $\bm{B}$ and $\bm{D}$ behave as 3-vector densities, while $\bm{E}$ and $\bm{H}$ behave as 3-covectors under the spatial coordinate transformations. This explains the position (upper vs lower) of indices in (\ref{EB}) and (\ref{DH}). 

In order to make electrodynamics a predictive theory, the system of differential equations (\ref{J1}) should be complemented by a constitutive relation between the electromagnetic excitation density ${\mathcal H}^{\mu \nu}$ and Maxwell's tensor $F_{\mu \nu}$. Although in general this may be a nonlinear and even nonlocal relation \cite{Hehl:2001,HeOb}, a wide class of phenomena, including gravitational and inertial effects, is described by the local linear constitutive relation. For a homogeneous and isotropic medium, characterized by the permittivity $\varepsilon$ and permeability $\mu$, moving in a curved manifold with the spacetime metric $g_{\mu\nu}$, such a constitutive relation reads \cite{ADP,TJP}:
\begin{equation}\label{HF}
{\mathcal H}^{\alpha\beta} = \lambda\,\sqrt{-g_{\rm opt}}\,g_{\rm opt}^{\alpha\mu}\,g_{\rm opt}^{\beta\nu}
\,F_{\mu\nu}\,,
\end{equation}
where 
\begin{equation}\label{gopt}
g_{\rm opt}^{\alpha\beta} = g^{\alpha\beta} + \left(1 - \varepsilon\mu\right)
{\frac {u^\alpha\,u^\beta}{c^2}}
\end{equation}
is the so-called {\it optical metric} that was first introduced by Gordon \cite{wgordon}, and the constant 
\begin{equation}
\lambda = \sqrt{\frac {\varepsilon\varepsilon_0}{\mu\mu_0}}\label{lam}
\end{equation}
has the dimensions of inverse impedance, namely, [ohm$^{-1}$]. Here, $\varepsilon$ and $\mu$ are the \emph{relative} permittivity and permeability of matter, respectively, and, as usual, $\varepsilon_0$ and $\mu_0$ are the dielectric and magnetic constants of the vacuum such that $\varepsilon_0\mu_0 = 1/c^2$. The motion of the medium is described by the average 4-velocity $u^\mu$, which is normalized as $g_{\mu\nu}u^\mu u^\nu = -\,c^2$. The covariant components of the optical metric (\ref{gopt}) read 
\begin{equation}\label{goptI}
g^{\rm opt}_{\alpha\beta} = g_{\alpha\beta} + \left(1 - {\frac 1{\varepsilon\mu}}\right)
{\frac {u_\alpha\,u_\beta}{c^2}}\,,
\end{equation}
and one straightforwardly verifies that $\sqrt{-g^{\rm opt}} = \sqrt{-\det g^{\rm opt}_{\alpha\beta}} = \sqrt{-g}\,/\sqrt{\varepsilon\mu}$. 

In terms of the 3-dimensional fields (\ref{EB}) and (\ref{DH}), one can recast (\ref{HF}) into a form of the constitutive relations of an anisotropic and inhomogeneous effective medium \cite{Sk,Pl,Fe,VIS,Bini:2018iyu}
\begin{equation}\label{J3}
D^a = \lambda\,\xi^{ab}E_b - (\bm{G} \times \bm{H})^a\,, \qquad
B^a = {\frac 1\lambda}\,\xi^{ab}H_b + (\bm{G} \times \bm{E})^a\,.
\end{equation} 
Here, the electric permittivity tensor density $\epsilon^{ab} = \lambda\,\xi^{ab}$, the magnetic permeability tensor density $\mu^{ab} = {\frac 1\lambda}\,\xi^{ab}$ and the gyration vector $\bm{G}$ of the effective optical medium are constructed in terms of the spacetime metric as
\begin{equation}\label{J4}
\xi^{ab} = -\,\sqrt{-g^{\rm opt}}\,\frac{g_{\rm opt}^{ab}}{g^{\rm opt}_{00}}\,,
\qquad G_a = - \,\frac{g^{\rm opt}_{0a}}{g^{\rm opt}_{00}}\,.
\end{equation}
These are conformally invariant features of the optical medium that supplants the background field, since source-free Maxwell's equations are conformally invariant. The influence of the gravity and inertia is encoded in the components of the spacetime metric. The latter, in general, can be conveniently described by the Arnowitt-Deser-Misner (ADM) parametrization \cite{ADM}
\begin{equation}\label{LT}
ds^2 = -\,V^2c^2dt^2 + \underline{g}{}_{ab}\,(dx^a - K^acdt)\,(dx^b - K^bcdt)\,,
\end{equation}
in terms of 10 functions $V = V(x^\mu)$, $K^a = K^a(x^\mu)$ and $\underline{g}{}_{ab}(x^\mu)$ that may depend arbitrarily on the local time and space coordinates $x^\mu = (t,x^a)$. In addition, let us write the components of the 4-velocity as
\begin{equation}\label{4U}
u^\mu = \gamma\left(1, \bm{v}\right)\,,
\end{equation}
where the normalization condition $g_{\mu\nu}u^\mu u^\nu = -\,c^2$ fixes the Lorentz factor as
\begin{equation}\label{gam}
  \gamma^{-2} = V^2 -
  \underline{g}{}_{ab}\,{\frac {\left(v^a - cK^a\right)\left(v^b - cK^b\right)}{c^2}}\,. 
\end{equation}

In the case of Minkowski spacetime $g_{\mu \nu} = \eta_{\mu \nu}$, we have $V = 1$, $\underline{g}{}_{ab} = \delta_{ab}$, $K^a = 0$, and hence one recovers $\epsilon^{ab} = \varepsilon\varepsilon_0\,\delta^{ab}$, $\mu^{ab} = \mu\mu_0\,\delta^{ab}$ and $\bm{G} = 0$, as expected. 

\subsection{Generalized Riemann-Silberstein vectors}

The electromagnetic field equations under consideration are linear; therefore, it is convenient to employ complex fields with the proviso that only their real parts have physical significance.  In terms of complex fields, we introduce the generalized Riemann-Silberstein \cite{RS1,RS2,Hehl:2004} vectors
\begin{equation}\label{J5}
\bm{\mathcal F}^{\pm} = \lambda\,\bm{E} \pm i\,\bm{H}\,, \qquad
\bm{\mathcal Z}_{\pm} = \bm{D} \pm i\,\lambda\,\bm{B}\,.
\end{equation} 
From the point of view of physics, both objects (\ref{J5}) look quite artificial, as the right-hand sides combine fields of different physical nature, i.e., electric and magnetic field strengths $(\bm{E}, \bm{B})$ are paired with the electric and magnetic excitations $(\bm{D}, \bm{H})$. However, from the geometric point of view, the 3-covector $\bm{\mathcal F}^{\pm}_a$ arises as a natural combination of two 3-covector fields, and similarly the 3-vector density $\bm{\mathcal Z}_{\pm}^a$ is naturally constructed from two 3-vector densities. As a result, Maxwell's equations in the optical medium can be written as~\cite{BM}
\begin{equation}\label{J6}
\bm{\nabla} \times \bm{\mathcal F}^{\pm} = \pm\,i\,\frac{\partial\bm{\mathcal Z}_{\pm}}{\partial t}\,,
\qquad \bm{\nabla}\cdot\bm{\mathcal Z}_{\pm} = 0\,,
\end{equation}
where the constitutive relations (\ref{J3}) are recast into
\begin{equation}\label{J7}
\bm{\mathcal Z}_{\pm}^a = \xi^{ab}\,\bm{\mathcal F}^{\pm}_b \pm i\,(\bm{G} \times \bm{\mathcal F}^{\pm})^a\,. 
\end{equation} 
In our linear perturbation treatment, the Dirac-type equation for the photon completely decouples for the $\bm{\mathcal F}^{+}$ and $\bm{\mathcal F}^{-}$ fields. Furthermore, Eq.~(\ref{J6}) implies $\partial_t(\bm{\nabla}\cdot\bm{\mathcal Z}_{\pm}) = 0$, which means that once $\bm{\nabla}\cdot\bm{\mathcal Z}_{\pm} = 0$ at any given time, then it is valid for all time. 

Henceforward, we limit our considerations to a stationary background and consider the propagation of electromagnetic radiation of frequency $\omega$ in the time-independent optical medium. The frequency of the radiation remains constant due to invariance under time translation. Let the fields depend upon time as
\begin{equation}\label{J8}
\bm{\mathcal F}^\pm = e^{-i\omega t}\,\bm{F}^\pm, \qquad
\bm{\mathcal Z}_{\pm} = e^{-i\omega t}\,\bm{Z}_\pm\,,
\end{equation} 
so that we can write Eq.~(\ref{J6}) for the time-independent factors as 
\begin{equation}\label{Ja}
\bm{\nabla}\times\bm{F}^{\pm} = \pm\,\omega\,\bm{Z}_{\pm}\,.
\end{equation} 
Then, $\bm{\nabla}\cdot\bm{Z}_{\pm} = 0$ is now automatically satisfied and the field equations (\ref{J6}) and (\ref{J7}) reduce to  
\begin{equation}\label{J9}
(\tfrac{1}{i}\,\bm{\nabla} - \omega\,\bm{G})\times\bm{F}^{\pm} = \mp\,i\,\omega\,\hat{\xi}\,\bm{F}^{\pm}\,,
\end{equation} 
where $\hat{\xi}$ denotes the linear algebraic map determined by the $3\times 3$ matrix $\xi^{ab}$.

Regarding the physical interpretation of our approach, let us consider a global inertial frame in Minkowski spacetime with inertial coordinates $(t_0, x_0, y_0, z_0)$ and metric
\begin{equation}\label{J10}
ds^2 = - \,c^2dt_0^2 + dx_0^2 + dy_0^2  + dz_0^2\,.
\end{equation}
In this inertial frame, a plane monochromatic electromagnetic wave of frequency $\omega_0$ and definite helicity propagates along the $z_0$ axis in vacuum with $\varepsilon = \mu = 1$ (therefore, $\lambda = \lambda_0 = \sqrt{\frac {\varepsilon_0}{\mu_0}}$ is the inverse of the impedance of free space). Let us write the corresponding  electric and magnetic fields as
\begin{equation}\label{J11}
\bm{E}_{\pm} = a_{\pm}\,(\bm{e}_{x_0}\pm i \bm{e}_{y_0})\,e^{-i\omega_0\,(t_0 -z_0/c)}\,, \qquad
\bm{B}_{\pm} = \frac{\mp\,i\,a_{\pm}}{c}\,(\bm{e}_{x_0}\pm i \bm{e}_{y_0})\,e^{-i\omega_0\,(t_0 -z_0/c)}\,,
\end{equation}
respectively. Here, $a_{+}$ ($a_{-}$) is a constant complex amplitude for positive (negative) helicity radiation. Accordingly, $\bm{H}_\pm = {\frac 1 {\mu_0}}\bm{B}_\pm$.

The Riemann-Silberstein vectors in this case, $\bm{\mathcal F}^{\pm}$, are such that for \emph{positive-helicity} radiation,  
\begin{equation}\label{J12}
\bm{\mathcal F}^{+} = \lambda_0\bm{E}_{+} + i\bm{H}_{+} = 2\lambda_0\,a_{+}\,(\bm{e}_{x_0} + i\bm{e}_{y_0})
\,e^{-i\omega_0\,(t_0 -z_0/c)}\,, \quad \bm{\mathcal F}^{-} = \lambda_0\bm{E}_{+} - i\bm{H}_{+}  = 0\,,
\end{equation} 
while for \emph{negative-helicity} radiation, 
\begin{equation}\label{J13}
\bm{\mathcal F}^{+} = \lambda_0\bm{E}_{-} + i\bm{H}_{-}  = 0\,, \quad \bm{\mathcal F}^{-} =
\lambda_0\bm{E}_{-} - i\bm{H}_{-} = 2\,\lambda_0a_{-}\,(\bm{e}_{x_0} - i\bm{e}_{y_0})
\,e^{-i\omega_0\,(t_0 -z_0/c)}\,.
\end{equation}
Moreover, we have
\begin{equation}\label{J14}
\bm{\nabla}\times\bm{\mathcal F}^{\pm} = \pm\,{\frac {\omega_0}c}\,\bm{\mathcal F}^{\pm}\,.
\end{equation} 
Thus in terms of complex fields, $\lambda_0\bm{E} + i\bm{H}$ essentially represents an electromagnetic wave with positive helicity, while $\lambda_0\bm{E} - i\bm{H}$ essentially represents a wave with negative helicity. 

It follows from these considerations that in our linear perturbation scheme, helicity is conserved in a scattering situation involving asymptotically flat spacetime. Moreover, $\bm{\mathcal F}^{+}$ and $\bm{\mathcal F}^{-}$ may be interpreted as helicity amplitudes in an arbitrary optical medium as a natural extension of the standard treatment in Minkowski spacetime.

We are interested in the implications of  the approach adopted here for the solution of Maxwell's equations in a uniformly rotating frame of reference.  

\subsection{Uniformly rotating system}

The spacetime metric in the rotating frame can be obtained from Eq.~(\ref{J10}) using the standard coordinate transformation~\cite{L+L}
\begin{equation}\label{J15}
t = t_0\,, \quad x = x_0 \,\cos \Omega t + y_0\, \sin \Omega t\,,\quad y = -x_0 \,\sin \Omega t + y_0\, \cos \Omega t\,, \quad z = z_0\,
\end{equation}
and the result is the special case of Eq.~(\ref{LT}) with $V = 1$ and $\underline{g}{}_{ab} = \delta_{ab}$:
\begin{equation}\label{J16}
ds^2 = - \left(1 - K^2\right) c^2dt^2 - 2 cdt\,(\bm{K}\cdot d\bm{r}) + dx^2 + dy^2  + dz^2\,,
\qquad \bm{K} = -\,{\frac 1c}\,\bm{\Omega}\times\bm{r}\,,
\end{equation}
where $\bm{r} = (x, y, z)$, $\bm{\Omega}$ is the rotation angular velocity and $K = |\bm{K}|$. 

Alternatively, we can focus on the fiducial observer that is spatially at rest at the origin of  inertial coordinates and refers its measurements to uniformly rotating axes. The natural orthonormal tetrad frame $e^{\mu}{}_{\hat \alpha}$ adapted to the noninertial observer can be expressed in $(t_0, x_0, y_0, z_0)$ coordinates by
\begin{equation}\label{J17}
e^{\mu}{}_{\hat 0} = (1, 0, 0, 0)\,, \qquad e^{\mu}{}_{\hat 1} = (0, \cos\Omega t, \sin\Omega t, 0)\,,
\end{equation}
\begin{equation}\label{J18}
e^{\mu}{}_{\hat 2} = (0, -\sin\Omega t, \cos\Omega t, 0)\,,\qquad e^{\mu}{}_{\hat 3} = (0, 0, 0, 1)\,.
\end{equation}
A geodesic (``Fermi") coordinate system can be established along the world line of the noninertial observer on the basis of the tetrad frame $e^{\mu}{}_{\hat \alpha}$; indeed, the Fermi coordinates and the Fermi metric coincide with the $(t, x, y, z)$ coordinates in Eq.~(\ref{J15}) and the metric in Eq.~(\ref{J16}). These coordinates are admissible for $K < 1$, which is the region within a cylinder of radius $c/\Omega$;  for a  thorough discussion, see~\cite{BMB}. In the rotating $(t, x, y, z)$ system, the Fermi transported tetrad frame reads \cite{HehlNi}
\begin{equation}\label{HN}
\tilde{e}{}_{\hat{t}} = \partial_t + cK^a\partial_a\,,\qquad \tilde{e}{}_{\hat{a}} = \partial_a\,.
\end{equation}
Replacing the spacetime within the uniformly rotating frame with the corresponding optical medium at rest with respect to the tetrad (\ref{HN}), i.e., the comoving matter with the 4-velocity $u = \tilde{e}{}_{\hat{t}}$, and using (\ref{J16}), (\ref{gopt}), and (\ref{goptI}), we find the components of the optical metric:    
\begin{eqnarray}\label{gopt00}
g^{\rm opt}_{00} &=& -\,{\frac {c^2}{\varepsilon\mu}} + c^2\delta_{ab}\,K^aK^b\,,
\quad g^{\rm opt}_{a0} = -\,c\,\delta_{ab}\,K^b\,,\\
g_{\rm opt}^{ab} &=& \delta^{ab} - \varepsilon\mu\,K^aK^b\,,\quad 
\sqrt{-g^{\rm opt}} = {\frac {c}{\sqrt{\varepsilon\mu}}}\,.\label{goptab} 
\end{eqnarray}
There is an alternative way to derive these results: one can consider the flat Minkowski spacetime with a homogeneous and isotropic medium at rest that has an optical metric given by Eq.~(\ref{goptI}), and apply the coordinate transformation (\ref{J15}) to the optical line element
\begin{equation}\label{dsopt}
ds_{\rm opt}^2 = - \,{\frac {c^2}{\varepsilon\mu}}\,dt_0^2 + dx_0^2 + dy_0^2  + dz_0^2\,.
\end{equation}

Accordingly, Eq.~(\ref{J4}) yields
\begin{equation}\label{J19}
\xi^{ab} = \chi^2\,{\frac {\sqrt{\varepsilon\mu}}{c}}
\left[\begin{array} {ccc}
1- {\frac {\varepsilon\mu\,\Omega^2y^2}{c^2}} & {\frac {\varepsilon\mu\,\Omega^2\,x\,y}{c^2}} & 0\\
{\frac {\varepsilon\mu\,\Omega^2\,x\,y}{c^2}} & 1 - {\frac {\varepsilon\mu\,\Omega^2\,x^2}{c^2}} & 0\\
0 & 0 & 1 \end{array}\right]\,
\end{equation}
and 
\begin{equation}\label{J20}
\bm{G} = -\,\chi^2\,{\frac {\varepsilon\mu}{c}}\bm{K}\,,
\end{equation}
where 
\begin{equation}\label{J21}
\chi^{-2} = 1 - {\frac {\varepsilon\mu\,\Omega^2 (x^2 + y^2)}{c^2}}\,.
\end{equation}
For the case $\varepsilon = \mu = 1$, the field equation (\ref{J9}) and some of its solutions have been investigated in~\cite{Hauck:2003gy}.  The present work is devoted to the solution of Eq.~(\ref{J9}) for a nontrivial medium with $\varepsilon\neq 1$ and $\mu\neq 1$ when terms quadratic in the angular velocity $\Omega$ are neglected. 

\subsection{Solution of Eq.~(\ref{J9}) to linear order in $\Omega$}

To first order in $\Omega$, we find 
\begin{equation}\label{K1}
\epsilon^{ab} \approx \varepsilon\varepsilon_0\,\delta^{ab}\,,\qquad \mu^{ab} \approx
\mu\mu_0\,\delta^{ab}\,,\qquad \bm{G} \approx {\frac {\varepsilon\mu}{c^2}}\,\bm{\Omega}\times\bm{r}
\end{equation}
and the constitutive relation (\ref{J7}) reads
\begin{equation}\label{ZF}
\bm{\mathcal Z}_{\pm} = {\frac {\sqrt{\varepsilon\mu}}{c}}\,\bm{\mathcal F}^{\pm}
\pm {\frac {i\,\varepsilon\mu}{c^2}}\,(\bm{\Omega}\times\bm{r})\times\bm{\mathcal F}^{\pm}. 
\end{equation} 
The field equation (\ref{Ja}) in this slowly rotating case reduces to
\begin{equation}\label{K2}
\bm{\nabla}\times\bm{F}^{\pm} = \pm\,{\frac {\omega\sqrt{\varepsilon\mu}}{c}}\,\bm{F}^{\pm}
+ {\frac {i\omega\,\varepsilon\mu}{c^2}}\,(\bm{\Omega}\times\bm{r})\times\bm{F}^{\pm}. 
\end{equation}
We look for a solution of the form  
\begin{equation}\label{K3}
\bm{F}^{\pm} =  \eta_{\pm}\,(\bm{e}_x \pm i \bm{e}_y)\,e^{i\,k^{\pm}\,z}
+ \bm{e}_z\,\zeta^{\pm}(x,y)\,e^{i\,k^{\pm}\,z}\,,
\end{equation} 
where $\eta_{\pm}$ are nonzero constants. 

Using ansatz (\ref{K3}), we find
\begin{eqnarray}\label{K4}
\bm{\nabla}\times\bm{F}^{\pm} \mp\,{\frac {\omega\sqrt{\varepsilon\mu}}{c}}\,\bm{F}^{\pm} =
\left[\frac{\partial \zeta^{\pm}}{\partial y}\pm \eta_{\pm}\left(k^{\pm}
- {\frac {\omega\sqrt{\varepsilon\mu}}{c}}\right)\right]\bm{e}_x\,e^{i\,k^{\pm}\,z}\nonumber\\
- \,\left[\frac{\partial \zeta^{\pm}}{\partial x} - i\, \eta_{\pm}\left(k^{\pm}
- {\frac {\omega\sqrt{\varepsilon\mu}}{c}}\right)\right]\bm{e}_y\,e^{i\,k^{\pm}\,z}
\mp {\frac {\omega\sqrt{\varepsilon\mu}}{c}}\zeta^{\pm}\,\bm{e}_z\,e^{i\,k^{\pm}\,z}\,.\label{K4a}
\end{eqnarray}
On the other hand, we have
\begin{equation}\label{K5}
{\frac {i\,\omega\,\varepsilon\mu}{c^2}}\,(\bm{\Omega}\times\bm{r})\times\bm{F}^{\pm}
= {\frac {i\,\omega\,\varepsilon\mu\,\Omega}{c^2}}\left[x\,\zeta^{\pm}\,\bm{e}_x
+ y\,\zeta^{\pm}\,\bm{e}_y - \eta_{\pm}(x \pm iy)\,\bm{e}_z\right]\,e^{i\,k^{\pm}\,z}\,.
\end{equation}
Equating the right-hand sides of Eqs.~(\ref{K4}) and (\ref{K5}) results in 
\begin{eqnarray}\label{K6}
\frac{\partial \zeta^{\pm}}{\partial y}\pm \eta_{\pm}\left(k^{\pm}
- {\frac {\omega\sqrt{\varepsilon\mu}}{c}}\right) &=&
{\frac {i\,\omega\,\varepsilon\mu\,\Omega\,x}{c^2}}\,\zeta^{\pm}\,,\\
\frac{\partial \zeta^{\pm}}{\partial x} - i\, \eta_{\pm}\left(k^{\pm}
- {\frac {\omega\sqrt{\varepsilon\mu}}{c}}\right) &=&
{\frac {-i\,\omega\,\varepsilon\mu\,\Omega\,y}{c^2}}\,\zeta^{\pm}\,,\label{K7}
\end{eqnarray}
\begin{equation}\label{K8}
  \zeta^{\pm} =  \pm \,{\frac {i\,\sqrt{\varepsilon\mu}\,\Omega}{c}}\,\eta_{\pm}(x \pm iy)\,,
\end{equation}
assuming that $\omega \ne 0$. Next, employing the explicit expression (\ref{K8}) for $ \zeta^{\pm}$ in Eqs.~(\ref{K6}) and (\ref{K7}), we get the same result to linear order in $\Omega$, namely,
\begin{equation}\label{K9}
\left(k^{\pm} - {\frac {\omega\sqrt{\varepsilon\mu}}{c}} \mp
{\frac {\Omega\sqrt{\varepsilon\mu}}{c}}\right)\eta_{\pm} = \mathcal{O}(\Omega^2)\,.
\end{equation}
Since $\eta_{\pm} \ne 0$ by assumption, we find that to linear order in $\Omega$
\begin{equation}\label{K10}
{\frac {c}{\sqrt{\varepsilon\mu}}}\,k^{\pm} = \omega \pm \Omega\,.
\end{equation}

In vacuum ($\varepsilon = \mu = 1$), Eq.~(\ref{K10}) agrees with Eq.~(\ref{I6}) that directly results from the spin-rotation coupling in this case. Recalling that $n = \sqrt{\varepsilon\mu}$ is the refractive index of the ambient medium, the general non-vacuum result (\ref{K10}) can be recast into
\begin{equation}\label{K11}
c\,k^{\pm} = n\,(\omega \pm \Omega)\,.
\end{equation}

\section{Spin-gravity coupling}

Mass current generates a gravitomagnetic field according to linearized general relativity resulting in the spin-gravitomagnetic field coupling~\cite{DeOT, Bini:2021gdb, Mashhoon:2023idh}. The Gravity Probe B (GP-B) space experiment has measured the gravitomagnetic field of the Earth~\cite{Francis1, Francis2}. The simple analogy with electrodynamics has its roots in the general linear approximation of general relativity. Let us imagine a global background inertial frame with Cartesian coordinates $x^\mu = (ct,\bm{x})$ that is perturbed by the weak gravitational field of a uniformly rotating astronomical source of constant mass $M$ and angular momentum $\bm{J}$. The spacetime metric is thus $g_{\mu\nu} = \eta_{\mu\nu} + h_{\mu\nu}(x)$, where $h_{\mu\nu}(x)$ is the perturbing field that is treated to linear order and $\eta_{\mu\nu} = {\rm diag}(-1, 1, 1, 1)$ is the Minkowski metric tensor of the background spacetime. Einstein's field equations with vanishing cosmological constant
\begin{equation}\label{L1} 
R_{\mu\nu} - {\frac{1}{2}}\,g_{\mu\nu}R = {\frac{8\pi G_{\rm N}}{c^4}}\,T_{\mu\nu}\,
\end{equation}
for the trace-reversed potentials $\bar{h}_{\mu\nu}=h_{\mu\nu}-\frac{1}{2}\eta_{\mu\nu}h$, $h = {\rm tr}(h_{\mu\nu})$, take the form
\begin{equation}\label{L2}
 \Box \bar{h}_{\mu\nu} = - \,{\frac{16\pi G_{\rm N}}{c^4}}\,T_{\mu\nu}\,,
\end{equation}
once the transverse gauge condition $\bar{h}^{\mu\nu}{}_{,\nu}=0$ has been imposed.  Here, $G_{\rm N}$ is Newton's constant of gravitation. We are interested in the special retarded solution of Eq.~(\ref{L2}) given by
\begin{equation}\label{L3} 
\bar{h}_{\mu\nu} = \frac{4G_{\rm N}}{c^4}\int\frac{T_{\mu\nu}
(ct-|\bm{x} - \bm{x}'|,\bm{x}')}{|\bm{x} - \bm{x}'|}\,d^3x'\,.
\end{equation} 

Regarding the properties of the astronomical source, we assume $T^{00}=\rho \,c^2$ and $T^{0a}=c\,j^a = c\,\rho \,v^a$, where $\rho(\bm{x})$ is the matter density and $\bm{v}$ is the velocity of the slowly moving matter, i.e., $|\bm{v}| \ll c$.  Moreover, $T_{ab}\sim \rho\, v_av_b+p\,\delta_{ab}$, where $p(\bm{x})$ is the pressure. It follows from Eq.~(\ref{L3}) that $\bar{h}_{ab} = {\mathcal O}(c^{-4})$. Henceforth, all terms of ${\mathcal O}(c^{-4})$ will be neglected in our treatment. Let us define the time-independent gravitoelectric ($\Phi_g$) and gravitomagnetic ($\bm{A}_g$) potentials as
\begin{equation}\label{L4} 
\bar{h}^{0 0} := \frac{4}{c^2}\, \Phi_g \,, \qquad  \bar{h}^{0a} := \frac{2}{c^2}\, A^a_g\,.
\end{equation} 
Then, the resulting spacetime metric is stationary and of the gravitoelectromagnetic (GEM) form~\cite{Bini:2021gdb, Mashhoon:2023idh}
\begin{equation}\label{L5} 
ds^2 = -\,\left(1-2\frac{\Phi_g}{c^2}\right)\,c^2\,dt^2-\frac{4}{c}(\bm{A}_g
\cdot d \bm{x})dt + \left(1+2\frac{\Phi_g}{c^2}\right) \delta_{ab}dx^adx^b\,.
\end{equation}

We define the stationary GEM fields in complete analogy with time-independent electrodynamics~\cite{Mashhoon:2000he}
\begin{equation}\label{L6} 
\bm{E}_g = -\,\bm{\nabla}\Phi_g\,, \qquad \bm{B}_g = \bm{\nabla} \times \bm{A}_g\,.
\end{equation}
In the exterior of the astronomical source, we have 
\begin{equation}\label{L7}
 \Phi_g \sim \frac{G_{\rm N}M}{|\bm{x}|}\,,\qquad \bm{A}_g \sim \frac{G_{\rm N}}{c}\frac{\bm{J}\times \bm{x}}{|\bm{x}|^3}\,.
\end{equation}
Let us note that the gravitoelectric field in this stationary case is essentially Newtonian, since $-\Phi_g$ reduces to the Newtonian gravitational  potential. On the other hand, the non-Newtonian gravitomagnetic field is given by 
\begin{equation}\label{L8}
\bm{B}_g \sim \frac{G_{\rm N}}{c\,|\bm{x}|^5}\, [\,3 (\bm{J} \cdot \bm{x})\, \bm{x} - \bm{J}\,|\bm{x}|^2\,]\,.
\end{equation}

It turns out that the gravitomagnetic field is locally equivalent to a rotation via the gravitational Larmor theorem, which is essentially the rotational part of Einstein's heuristic principle of equivalence~\cite{Larmor, Bahram}. Let us consider a free test gyroscope with its center of mass held at rest in a gravitational field. The local gravitomagnetic field $\bm{B}_g(\bm{x})$  causes a precession of the gyroscope's spin vector with a precession frequency given by $\bm{B}_g/c$. Let us now imagine supplanting the gravitomagnetic field by a rotating frame in the neighborhood of the gyroscope. The observed motion of the gyroscope would be the same as before if observers at rest in the rotating frame rotate with Larmor frequency~\cite{Bahram}  
\begin{equation}\label{L9}
\bm{\Omega}_{\rm L} = -\,{\frac{1}{c}}\,\bm{B}_g\,.
\end{equation}
 It follows from Eq.~(\ref{I1}) that the resulting Hamiltonian for the intrinsic spin-gravity coupling is of the form~\cite{Papini:2007gx, Mashhoon:2013jaa, Mashhoon:2023idh} 
\begin{equation}\label{L10}
\mathcal{H}_{sg} = \frac{1}{c} \,\bm{S} \cdot \bm{B}_g\,.
\end{equation}
The possibility of measuring the intrinsic spin-gravity coupling has been considered by a number of investigators;  see, for instance~\cite{Bah1, Bah2, Tarallo:2014oaa, Fadeev:2020gjk, Vergeles:2022mqu} and the references cited therein.

In principle, the spin-of-light gyroscope of Fedderke \emph{et al.}~\cite{Fedderke:2024ncj} can be employed to measure the component of the local gravitomagnetic field along the principal gyroscope axis. Consider, for the sake of illustration, the propagation of electromagnetic waves along the axis of rotation of the astronomical body under consideration here. Indeed, we want to measure the gravitomagnetic field at some point along the rotation axis.  Let the angular momentum of the source be along the $z$ direction and the electromagnetic waves propagate along the $z$ axis away from the source; that is,
\begin{equation}\label{L11}
\bm{J} = J\, \bm{e}_{z}\,, \qquad  |\bm{B}_g| = \frac{2 G_{\rm N}}{c}\,\frac{J}{z^3}\,,
\end{equation}
where $z$ here denotes a point in the exterior of the source on the $z$ axis.  An analysis of the spin-gravity coupling in this case results in an equation analogous to Eq.~(\ref{I5}), namely, 
\begin{equation}\label{L12}
\omega = c\,k^{\pm}(\bm{x}) \pm \frac{1}{c}\,\hat{\bm{k}}^{\pm}(\bm{x}) \cdot \bm{B}_g(\bm{x})\,,
\end{equation}
which in the case of propagation along the axis of rotation implies
\begin{equation}\label{L13}
c\,k^{\pm}(z) =  \omega \mp \frac{2 G_{\rm N}}{c^2}\,\frac{J}{z^3}\,.
\end{equation}
Clearly, the spin-of-light gyroscope of Fedderke \emph{et al.}~\cite{Fedderke:2024ncj} can be employed in principle to measure local gravitomagnetic fields; however, the gravitational effects are indeed rather weak. For instance, for the angular velocity of the Earth, $\Omega_{\oplus}/(2\pi) \approx 11.6\, \mu$Hz, we have $\hbar \Omega_{\oplus} \approx 5 \times 10^{-20}$ eV, while $\hbar |\bm{B}_g|/c  \approx 10^{-29}$ eV, where in Eq.~(\ref{L11}) we take $z = R_{\oplus}$ to be the radius of the Earth. 

Apart from the significance of dispersion relation (\ref{L12}) for the possibility of measurement of the gravitomagnetic field via the the spin-of-light gyroscope, this relation is important for the propagation of radiation in the exterior gravitational field of a rotating source. This is demonstrated in Appendix~\ref{appB}, where Eq.~(\ref{L12}) is employed to describe the gravitational Faraday rotation of the plane of linear polarization in the gravitomagnetic field of a uniformly rotating mass. 

\section{Discussion} 

We study the implications of spin-rotation coupling in connection with a new gyroscope recently proposed by Fedderke \emph{et al.}~\cite{Fedderke:2024ncj}. In this connection, the propagation of electromagnetic radiation in a system that uniformly rotates about the direction of wave propagation is investigated in the presence of an ambient medium. Moreover, we discuss the possibility of using this type of spin-of-light gyroscope to measure gravitomagnetic fields. 

Should highly precise spin-of-light gyroscopes become available in the future, then one may consider the possibility of exploring the detailed nature of solar and planetary gravitomagnetic fields via spacecrafts on general trajectories within the solar system. In particular, it may become possible to measure the temporal variation of the gravitomagnetic field that could be due to the intrinsic time dependence of the gravitational source~\cite{Mashhoon:2008kq, Iorio:2024ell}. Finally, we note that the spacetime curvature associated with a gravitational radiation field has both gravitoelectric and gravitomagnetic components leading to gravitoelectric and gravitomagnetic fields, respectively,  in close analogy with electrodynamics~\cite{MLR, Mashhoon:2021qtc}. The gravitoelectric fields have been employed thus far to detect gravitational waves via the Jacobi equation (i.e., the geodesic deviation equation). It may become feasible in the future to measure the corresponding gravitomagnetic fields as well; however, in this endeavor, sufficiently accurate gyroscopes will be indispensable~\cite{MaTh}.


\appendix

\section{Rotational ``Doppler" Effect}\label{appA}

Imagine a plane monochromatic electromagnetic radiation of frequency $\omega_0$, wave vector $\bm{k}_0 =  k_0\,\bm{e}_{z_0}$ and definite helicity propagating along the $z_0$ axis of a global inertial frame. Static observers are at rest all along the $z_0$ axis.  These static observers are not inertial, since they refer their measurements to Cartesian axes that rotate with constant angular velocity $\Omega$ about the $z_0$ axis. In the inertial frame, $F_{\mu \nu} \mapsto (\bm{E}_{\pm}, \bm{B}_{\pm})$, where the corresponding electric and magnetic fields are given by Eq.~(\ref{J11}). The observers at rest on the axis measure the electromagnetic field tensor
\begin{equation}\label{A1}
F_{\hat \alpha \hat \beta} = F_{\mu \nu}\, e^{\mu}{}_{\hat \alpha}\, e^{\nu}{}_{\hat \beta}\,,
\end{equation}
which is the projection of the inertial electromagnetic field tensor on the tetrad frame field $e^{\mu}{}_{\hat \alpha}$ adapted to the noninertial static observers and given by Eqs.~(\ref{J17})--(\ref{J18}). Explicit calculations reveal that 
\begin{equation}\label{A2}
(F_{\hat 0 \hat 1}, F_{\hat 0 \hat 2}, F_{\hat 0 \hat 3}) = -\,a_{\pm}\,(1, \pm\,i, 0) \,e^{-i\,(\omega_0\,\mp\,\Omega)t_0  + i k_0z_0}\,, 
\end{equation}
\begin{equation}\label{A3}
(F_{\hat 2 \hat 3}, F_{\hat 3 \hat 1}, F_{\hat 1 \hat 2}) = \frac{1}{c}\, a_{\pm}\,(\mp\,i, 1, 0) \,e^{-i\,(\omega_0\,\mp\,\Omega)t_0  + i k_0z_0}\,,
\end{equation}
where $\omega_0 = c\,k_0$. 

It follows from the Fourier analysis of the measured electromagnetic field tensor that the frequency and wave vector of the radiation as measured by the noninertial observers all along the rotation axis are given by
\begin{equation}\label{A4}
\omega = \omega_0 \mp \Omega\,, 
\end{equation}
and 
\begin{equation}\label{A5}
\bm{k} = \bm{k}_0\,. 
\end{equation}
 
When positive (negative) helicity radiation propagates along the $z_0$ direction, the electric and magnetic fields rotate with angular speed $\omega$ in the positive (negative) sense about the direction of propagation. However, a noninertial observer at rest on the $z_0$ axis perceives the situation differently: For waves of positive (negative) helicity, the electric and magnetic fields appear to rotate with angular speed $\omega - \Omega$ ( $\omega + \Omega$ ) in the positive (negative) sense about the $z_0$ axis. This particular composition of frequencies is reminiscent of the composition of linear velocities in the Doppler effect~\cite{BMB, Mashhoon:2024qwj}. 

The dispersion relation for the noninertial observers can be simply worked out using Eqs.~(\ref{A4}) and (\ref{A5}). The spacetime of noninertial observers is stationary; hence,  $\omega$ is constant for the two helicity states. It then follows that $ck^{\pm} = \omega \pm \Omega$. 

\section{Gravitational Faraday Rotation}\label{appB}

The helicity dependence of the wave vector in the rotating frame is entirely analogous to the circumstance surrounding the propagation of electromagnetic radiation in the stationary spacetime of a rotating source that results in the Skrotskii effect, namely, the rotation of the plane of linear polarization also known as the gravitational Faraday effect. 

In the exterior of the uniformly rotating source,  
\begin{equation}\label{B1}
\bm{B}_g = - \,\bm{\nabla}\chi_g\,, \qquad \chi_g =
\frac{G_{\rm N}}{c}\frac{\bm{J} \cdot \bm{x}}{|\bm{x}|^3}\,,
\end{equation}
since $\bm{B}_g$  is curl free. Here,  $\chi_g$ is the gravitomagnetic scalar potential.  A  straightforward consequence of Eq.~(\ref{L13}) is that the plane of linear polarization  of an electromagnetic wave that starts at some initial point $z = z_i$ along the axis of rotation away from the source and propagates along the axis will rotate by an angle of~\cite{Ramos:2006sb} 
\begin{equation}\label{B2}
\Delta = \frac{1}{2}\,\int_{z_i}^{z} (k^{-} - k^{+}) dz = \frac{1}{c^2}\,
[\chi_g(z_i) - \chi_g(z)] = \frac{G_{\rm N}}{c^3}\,J \left(\frac{1}{z_i^2} - \frac{1}{z^2}\right)\,.
\end{equation}
This is the gravitational Faraday rotation first elucidated by Skrotskii~\cite{Sk}. See~\cite{Ramos:2006sb}  for details of the derivation and for extension to the case of gravitational radiation. For recent discussions of the effect, see~\cite{Shoom:2024zep} and the references cited therein.

\end{document}